\documentstyle[twocolumn,aps,prl,floats,ifthen,epsf]{revtex}

\newcommand{\showfigs}{true}

\begin{document}

\draft

 \wideabs{

\title{Parton energy loss limits and shadowing
in Drell-Yan dimuon production}

\author{
M.A.~Vasiliev$^i$\cite{byline1},
M.E.~Beddo$^g$,
C.N.~Brown$^c$,
T.A.~Carey$^f$,
T.H.~Chang$^g$,
W.E.~Cooper$^c$,
C.A.~Gagliardi$^i$,
G.T.~Garvey$^f$,
D.F.~Geesaman$^b$,
E.A.~Hawker$^{i,f}$,
X.C.~He$^d$,
L.D.~Isenhower$^a$,
D.M.~Kaplan$^e$,
S.B.~Kaufman$^b$,
D.D.~Koetke$^j$,
W.M.~Lee$^d$,
M.J.~Leitch$^f$,
P.L.~McGaughey$^f$,
J.M.~Moss$^f$,
B.A.~Mueller$^b$,
V.~Papavassiliou$^g$,
J.C.~Peng$^f$,
G.~Petitt$^d$,
P.E.~Reimer$^f$,
M.E.~Sadler$^a$,
W.E.~Sondheim$^f$,
P.W.~Stankus$^h$,
R.S.~Towell$^{a,f}$,
R.E.~Tribble$^i$,
J.C.~Webb$^g$,
J.L.~Willis$^a$,
G.R.~Young$^h$\\
(FNAL E866/NuSea Collaboration)
}

\address{
$^a$Abilene Christian University, Abilene, TX 79699\\
$^b$Argonne National Laboratory, Argonne, IL 60439\\
$^c$Fermi National Accelerator Laboratory, Batavia, IL 60510\\
$^d$Georgia State University, Atlanta, GA 30303\\
$^e$Illinois Institute of Technology, Chicago, IL  60616\\
$^f$Los Alamos National Laboratory, Los Alamos, NM 87545\\
$^g$New Mexico State University, Las Cruces, NM, 88003\\
$^h$Oak Ridge National Laboratory, Oak Ridge, TN 37831\\
$^i$Texas A \& M University, College Station, TX 77843\\
$^j$Valparaiso University, Valparaiso, IN 46383
}
\date{\today}

\maketitle

\begin{abstract}
A precise measurement of the ratios of the Drell-Yan cross
section per nucleon for an 800 GeV/$c$
proton beam incident on Be, Fe and W targets is
reported.
The behavior of the Drell-Yan ratios at small target parton
momentum fraction
is well described by an existing fit to the shadowing
observed in deep-inelastic scattering.
The cross section ratios as a
function of the incident parton momentum fraction
set tight limits on the energy loss of quarks
passing through a cold nucleus.
\end{abstract}


 } 

The Drell-Yan process, where a beam quark (antiquark) fuses with
a target
antiquark (quark) producing a muon pair, can be used to study
the
interactions of fast partons penetrating through cold nuclei.  Only
initial
state interactions are important in Drell-Yan since the dimuon
in the
final state does not interact strongly with the partons in the medium. 
This makes Drell-Yan scattering an ideal tool to study energy loss of
fast quarks in nuclear matter by comparing the observed yields
from a range of nuclear targets.  The dynamics of fast parton energy
loss in nuclear matter is the subject of considerable theoretical
interest \cite{gavin,brodsky,baier1,baier2}
and has significant implications for the physics of
relativisitic heavy ion collisions.

Drell-Yan scattering is closely related to deep-inelastic scattering
(DIS) of
leptons, but unlike DIS it can be used specifically to probe
antiquark
contributions in target parton distributions.  When DIS
on nuclei occurs at $x < 0.08$, where $x$
is the parton momentum fraction, the cross section
per nucleon decreases with increasing nuclear number $A$ due to
shadowing \cite{shadow1,shadow2}.  Shadowing should also
occur in Drell-Yan
dimuon production at small $x_2$, the momentum fraction of the
target parton, and theoretical calculations indicate that
shadowing in the two reactions has a common origin
\cite{Kope96,Brod98}.  This should be particularly apparent
at $x<0.06$ where DIS on nuclei, like Drell-Yan, is dominated
by scattering off sea quarks.

Fermilab Experiment 866 (E866) measured the nuclear dependence of
Drell-Yan dimuon production by 800 GeV/$c$ protons on Be, Fe
and W targets at larger
values of $x_1$, the momentum fraction of the beam parton,
larger values of $x_F$ ($\approx x_1 - x_2$), and
smaller values of $x_2$ than reached by the previous 
experiment, Fermilab E772 \cite{E772}.  The extended 
kinematic coverage of E866 significantly
increases its sensitivity
to parton energy loss and shadowing.  This Letter reports the
results.

\begin{figure*}[tb]
\ifthenelse{\equal{\showfigs}{true}}{
  \begin{center}
    \mbox{\epsfxsize=6.9in \epsfbox{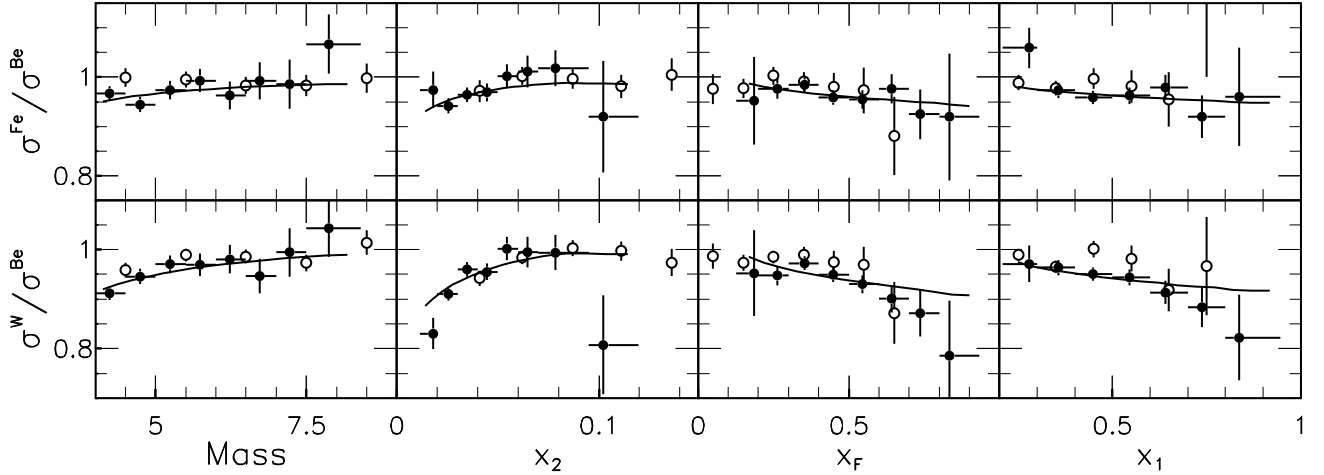}}
  \end{center}
}{}
  \caption{Ratios of the measured cross section per nucleon
for Drell-Yan events versus dimuon mass (in GeV/$c^2$),
$x_2$, $x_F$ and $x_1$. The upper (lower) panels
show ratios of Fe/Be (W/Be) from the present experiment as solid circles and
Fe/C (W/C) from E772 as open circles.  The errors are statistical only.
The solid curves are the predicted cross section ratios, integrated over
the hidden variables,
from leading-order calculations using
EKS98 \protect\cite{esk} and MRST \protect\cite{MRST}.}
\label{fig:crossratio}
\end{figure*}

The experiment used the same 3-dipole magnet
spectrometer that was described in \cite{hawker}.
An 800~GeV/$c$
extracted proton beam averaging $3 \times 10^{11}$ protons per
20~s spill
bombarded one of three solid targets or an empty target frame. 
The Be, Fe
and W targets were 9.4\% -- 19\% of an interaction length thick. 
Their relative thicknesses were chosen carefully
to match the background rates present in the spectrometer.  The
targets were located
far upstream of the main dipole magnet to
optimize the
acceptance at low $x_2$ and large $x_F$. 
During a cycle, 2 beam spills were taken on each
target and 1 spill was collected on an empty location.  After passing
through a target, the remaining beam was intercepted by a copper
beam dump, which was followed by a thick absorber that
removed hadrons produced in the target and the
dump.  This ensured that only muons traversed the spectrometer's
detectors, which consisted
of four tracking stations and a
momentum analyzing magnet.  
The trigger required a pair of triple hodoscope coincidences
having the topology of a muon pair from the
target.  Typically 1400 triggers per second were recorded with an
average live time of 93\%.

Over 130,000 muon pairs with dimuon mass in the range
$4.0 < M <
8.4$ GeV/$c^2$ survived the data cuts.  Random pairs were
subtracted from the data using simulated random
dimuons constructed by mixing single
muon tracks that were obtained simultaneously from
prescaled single muon triggers.  The normalization factor for the
random
correction was evaluated by comparing to the observed rate of
same-charge
muon pairs in the data. The random rate was 9\% of the real
events, and it occurred primarily at low dimuon
mass and high $p_T$.  A residual 3\% background from the
empty target was also subtracted.  The solid
targets intercepted the beam at nearly the same $z$ location,
making differences in dimuon acceptance
negligible.  The overall systematic normalization
uncertainty in the cross section
ratios reported here is 1\%.

The Drell-Yan events obtained by E866 extend over the ranges
$0.01<x_2<0.12$ and
$0.21<x_1<0.95$, with $\langle x_2 \rangle$ = 0.038 and
$\langle x_1 \rangle$ = 0.46.  They also
cover the range $0.13<x_F<0.93$ and provide good $p_T$
coverage to 4 GeV/$c$.  Ratios
of the cross section per nucleon for Fe to Be and W to Be versus
dimuon mass, $x_2$, $x_F$ and $x_1$
are shown in Fig.~\ref{fig:crossratio}, along with similar results
from E772 for Fe to C and W to C.  The figure shows very good
agreement between the experiments for the cross section 
ratios versus $x_2$.
The agreement versus other variables is also quite good, given
the much smaller $\langle x_2 \rangle$ and, hence, increased
shadowing of the present
data.  Note that the $A$ dependence observed here 
and in E772 implies
the change in the cross section ratios in
Fig.~\ref{fig:crossratio} due to the choice of Be
versus C is small compared to the effect of the difference in
$\langle x_2 \rangle$.
The reduction in the cross section per nucleon on
the heavy targets, characteristic of shadowing, is clearly apparent
at small $x_2$.  A similar
reduction, part of which could be related to incident parton energy
loss, is apparent at large
$x_F$ and $x_1$.  However, it is important to recognize that the
spectrometer acceptance coupled to
the intrinsic Drell-Yan cross section leads to a strong
anti-correlation between $x_2$ and $x_F$
for the observed events.  Therefore, the events that show the cross
section reduction at large
$x_F$ and $x_1$ are in general the same events that appear in the shadowing
region.

In order to identify the contributions from shadowing,
Fig.~\ref{fig:crossratio} also shows the predicted cross section
ratios, integrated over the
hidden variables, from leading-order Drell-Yan calculations
using the code EKS98 \cite{esk} together with the MRST
parton distribution functions \cite{MRST}.
Essentially identical results are
obtained using CTEQ5M or CTEQ5L parton
distributions \cite{CTEQ5} instead.
EKS98 describes the ratios $f_A(x,Q^2)/f_p(x,Q^2)$ of the
various quark flavors in nucleus $A$, compared to those in
the proton.  It has been tuned to fit the shadowing observed
in DIS \cite{NMC} and E772 while conserving baryon
number and momentum.

EKS98 provides an unbiased way to separate the effects of
shadowing and energy loss in the present data because it
uses a single shadowing function
to describe the nuclear dependence of all sea quark distributions
at its initial scale ${Q_0}^2$ = 2.25 GeV$^2$
and only DIS results
are used to constrain that function for $x \lesssim 0.1$.
The shape and magnitude of the ratios versus
$x_2$ are well reproduced by the shadowing
predictions.  Most of the
$A$ dependence observed in the ratios versus mass, $x_F$
and $x_1$ can also
be explained by shadowing at small $x_2$.
As a further test of the shadowing parametrization, the events in
Fig.~\ref{fig:crossratio} have been separated into two sets at
the median $x_2$ value.  The cross section ratios differ by 
up to 6\% when the events at low and
high $x_2$ -- but at the same mass, $x_F$ or $x_1$ -- are 
observed separately.  EKS98 also
describes these differences well.
This is the first experimental
demonstration that the shadowing
observed in Drell-Yan and DIS is quantitatively similar.

\begin{figure}[tb]
\ifthenelse{\equal{\showfigs}{true}}{
  \begin{center}
    \mbox{\epsfxsize=8cm \epsfbox{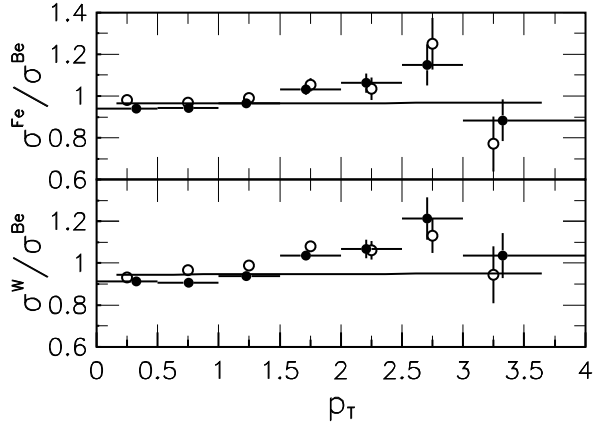}}
  \end{center}
}{}
  \caption{Ratios of the measured Drell-Yan cross section per
nucleon versus $p_T$.  Ratios of Fe/Be and W/Be from the present
experiment are shown as solid circles, and ratios of Fe/C and W/C
from E772 are shown as open circles.  The solid curves are
shadowing predictions for the present experiment from
leading-order calculations using
EKS98 \protect\cite{esk} and MRST \protect\cite{MRST}.}
\label{fig:pt}
\end{figure}

Figure \ref{fig:pt} shows the measured ratios of the Drell-Yan
cross section per nucleon as a function of
$p_T$.  Unlike the longitudinal variables that
have strongly correlated acceptances, the
$p_T$ acceptance in E866 is nearly independent of the other kinematic
variables, so the shadowing calculations predict essentially
constant cross section ratios versus $p_T$, as shown by the smooth
curves in the figure.  The data demonstrate a clear $p_T$ dependence
which must have an independent origin.
The slight reduction in the cross section per nucleon
observed for heavy nuclei at small
$p_T$, coupled with the increase in the cross section per nucleon
at large $p_T$, is characteristic
of multiple scattering of the incident partons as they traverse the
nucleus.

\begin{figure}[tb]
\ifthenelse{\equal{\showfigs}{true}}{
  \begin{center}
    \mbox{\epsfxsize=8cm \epsfbox{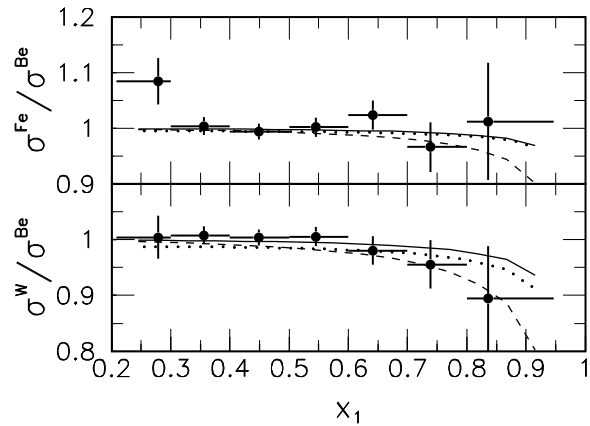}}
  \end{center}
}{}
  \caption{Ratios of the cross section per nucleon versus $x_1$ for
Fe/Be (upper panel) and W/Be (lower panel), corrected for shadowing.
The solid curves are the best fit using the energy loss form 
(\ref{eq:1}), and the dashed curves show the $1\sigma$ upper limits.
The dotted curves show the $1\sigma$ upper limits using the energy
loss form (\ref{eq:3}).  The $1\sigma$ upper
limit curves using the energy loss form (\ref{eq:2}) are essentially
identical to those using form (\ref{eq:3}).}
\label{fig:x1_ratio}
\end{figure}

The $x_1$ dependence of the cross section ratios provides the best
direct measure of the energy loss of the incident quarks in the
nuclear medium.  Table~\ref{tab:1} 
gives the ratios of the measured cross
section per nucleon as a function of $x_1$, and the mean values of
$x_1$, $x_2$ and dimuon mass for each bin.
However, as shown above, shadowing
at small $x_2$ explains a substantial fraction of the apparent variation
in the cross section ratios versus $x_1$.  This must be removed before
one can isolate a nuclear dependence due to energy loss.
Figure \ref{fig:x1_ratio}
shows the same cross section ratios
versus $x_1$ as given in Table~\ref{tab:1}, but corrected for
shadowing
by weighting
each event with the calculated ratio of the Drell-Yan
cross sections per nucleon for deuterium and
nucleus $A$ at the same $(x_1,x_2,Q^2)$,
using EKS98 and MRST. 

\begin{table}[tb]
  \caption{Ratios of the cross section per nucleon for
Fe to Be and W to Be 
as functions of $x_1$, without correction for shadowing.
Also given are mean values for $x_1$,
$x_2$ and
dimuon mass (in GeV/$c^2$) for each bin.
Mean $x_F$ may be found from
$\langle x_F \rangle = \langle x_1 \rangle -
\langle x_2 \rangle$.  
The errors are statistical only.}
  \label{tab:1}
  \begin{tabular}{cccccc}
$x_1$ range & $\langle x_1 \rangle$  & $\langle x_2 \rangle$ &
$\langle M \rangle$ & $\sigma^{\rm Fe}/\sigma^{\rm Be}$ & 
 $\sigma^{\rm W}/\sigma^{\rm Be}$\\ \hline
0.21-0.3& 0.279 & 0.055 & 4.76 & 1.059(41) & 0.971(37) \\ 
0.3-0.4 & 0.356 & 0.046 & 4.92 & 0.973(16) & 0.964(15) \\ 
0.4-0.5 & 0.448 & 0.039 & 5.07 & 0.959(14) & 0.951(13) \\ 
0.5-0.6 & 0.545 & 0.034 & 5.17 & 0.962(17) & 0.944(16) \\ 
0.6-0.7 & 0.642 & 0.030 & 5.24 & 0.979(25) & 0.914(23) \\ 
0.7-0.8 & 0.738 & 0.026 & 5.27 & 0.920(42) & 0.884(40) \\ 
0.8-0.95& 0.836 & 0.023 & 5.28 & 0.960(99) & 0.822(86) \\ 
  \end{tabular}
\end{table}

Several groups have studied
energy loss of partons in nuclei.  Their results can be
expressed in
terms of the average change in the incident parton
momentum fraction, ${\Delta}x_1$, as a function of target nucleus. 
Gavin and Milana~\cite{gavin} analyzed the E772 Drell-Yan data
for energy loss based on the parametrization 
\begin{equation} 
{\Delta}x_1 = -{\kappa}_1\,x_1\,A^{1/3},
\label{eq:1}
\end{equation}
where the factor ${\kappa}_1$ may have a $Q^2$ dependence. 
They based the
form (\ref{eq:1}) on an analogy to
the transverse spin asymmetry in direct photon production.
From a comparison to the $x_F$ dependence
seen by E772 and neglecting shadowing,
they concluded that the fractional energy loss of quarks
passing through nuclei is $\approx$\,0.4\%/fm.  Recently,
other groups have assumed equivalent formulations
with even more rapid energy loss \cite{Gale99,Marc99}. 
Brodsky and Hoyer~\cite{brodsky} argued
that the energy
loss found by Gavin and Milana was too large since the time scale
for QCD
bremsstrahlung was too short to allow for multiple contributions to
the energy loss.  Brodsky and Hoyer used an analogy to the 
photon bremsstrahlung process to obtain a form for gluon
radiation leading to an initial parton energy loss \cite{Brod99}
\begin{equation} 
{\Delta}x_1 \approx -\frac{{\kappa}_2}{s} A^{1/3}.
\label{eq:2}
\end{equation}
They found an upper limit for the gluon radiation
from the uncertainty
relation.  They also noted that elastic scattering should make
a similar contribution to the energy loss.  Overall,
they concluded that energetic partons should lose
$\lesssim$\,0.5 GeV/fm in nuclei.
The formulation developed by Brodsky and Hoyer was
extended by Baier {\it et al}.~\cite{baier1,baier2}.
They found that the energy loss
of sufficiently energetic partons depends on a
characteristic length and the broadening of the squared transverse
momentum of the parton.  For finite nuclei, both factors vary
as $A^{1/3}$, so Baier {\it et al}.\@ predict 
\begin{equation} 
{\Delta}x_1 \approx -\frac{{\kappa}_3}{s} A^{2/3}.
\label{eq:3}
\end{equation}
Baier {\it et al}.\@ predict that energy loss may be
different in hot and cold nuclear matter and that
there could a large coherent effect in
relativistic heavy ion collisions.  In contrast, tight limits have
been placed on energy loss in S+S and Pb+Pb collisions
\cite{Wang98}, but the parton momenta relative to the
hadronic medium were very much smaller in those
cases than in the present experiment, making a 
direct comparison difficult.

Given these energy loss expressions, it is possible to obtain
empirical
values for the $\kappa$'s by performing simultaneous fits to the
Fe/Be and W/Be Drell-Yan cross section
ratios versus
$x_1$ in Fig.~\ref{fig:x1_ratio}.
We assume that ${\kappa}_1$ is $Q^2$-independent
\cite{gavin,brodsky} when using form (\ref{eq:1}) and that our
incident quarks are sufficiently energetic when using form
(\ref{eq:3}).  Curves corresponding to the fits are
included in Fig.~\ref{fig:x1_ratio}.
When assuming the form (\ref{eq:1}), we find 
$\kappa_1 = 0.0004 \pm 0.0009$.
This implies that the observed fractional energy loss of the
incident quarks is $<$\,0.14\%/fm.
For the energy loss forms (\ref{eq:2}) and
(\ref{eq:3}), the best fits imply essentially zero energy loss.  We
find the $1\sigma$ upper limits to be $\kappa_2 < 0.75$ GeV$^2$ and
$\kappa_3 < 0.10$ GeV$^2$.  The $\kappa_2$ limit indicates
that the incident quarks lose energy at a constant rate of
$<$\,0.44 GeV/fm.
The $\kappa_3$ limit implies that the observed energy loss
of the incident quarks within the model of
Baier {\it et al}.\@ is
$\Delta E < 0.046$ GeV/fm$^2$ $\times$ $L^2$, where
$L$ is the quark propagation length through the nucleus.
This is very close to
the lower value given by Baier {\it et al}.\@ for
cold nuclear matter \cite{baier3}.
In all three cases, the quoted errors include both
statistics and the overall normalization uncertainty,
with the latter dominating.

One can also obtain an indirect
estimate of the energy loss due to gluon radiation
in the model of Baier {\it et al}.\@
from the broadening of the $\langle {p_T}^2 \rangle$
as the incident quark passes through the nucleus,
as shown in Fig.~\ref{fig:pt}.
However, such an analysis involves two significant
difficulties.  While the Drell-Yan cross section is
very small for $p_T$ beyond 4 GeV/$c$, the change in
$\langle {p_T}^2 \rangle$ from nucleus to nucleus,
being the small difference of large numbers, is
quite sensitive to the yield at very large $p_T$
where our acceptance is poor and the random
background becomes significant.  Meanwhile, the $p_T$
dependence of the ratio of the Drell-Yan cross sections
per nucleon on hydrogen and deuterium shows possible
evidence for a change in the reaction mechanism at
$p_T \approx 3$ GeV/$c$ \cite{rst99}, complicating
interpretation of the large $p_T$ ratio data 
in Fig.~\ref{fig:pt}.  Further
analysis of the $p_T$ dependence of the
Drell-Yan cross section will be presented
in a future publication.

In summary, this Letter reports a measurement of the ratios
of the Drell-Yan cross
section per nucleon for Fe to Be and W to Be.  Nuclear shadowing
is
found to be important in the small $x_2$ domain.  For sufficiently
small $x_2$, the
shadowing observed in Drell-Yan
has been demonstrated to be quantitatively similar to that in DIS. 
Subsequently, a
correction for shadowing has been applied.
The cross section ratios versus $x_1$ 
provide direct upper limits
on the energy loss of the incoming parton as it traverses a cold
nucleus that are tighter than previous constraints. 
Shadowing and initial state energy loss are processes that occur
in both Drell-Yan production and $J/\psi$
formation.  Hence, these results should also further the understanding of
$J/\psi$ production, which is
required if it is to be used as a signal for the quark-gluon plasma
in relativistic heavy ion collisions.

We would like to acknowledge R. Vogt for many useful discussions.
We thank the Fermilab Particle Physics, Beams and
Computing Divisions for their assistance in performing this
experiment.  This work was supported in part by the U.S.
Department of Energy.


\begin{references}

\bibitem[*]{byline1}On leave from Kurchatov Institute,
Moscow, Russia.
\bibitem{gavin} S. Gavin and J. Milana, Phys. Rev. Lett. {\bf 68},
1834 (1992).
\bibitem{brodsky} S.J. Brodsky and P. Hoyer, Phys. Lett. B {\bf
298}, 165 (1993).
\bibitem{baier1} R. Baier {\em et al.}, Nucl. Phys. B {\bf 484},
265 (1997).
\bibitem{baier2} R. Baier {\it et al}., hep-ph/9804212.
\bibitem{shadow1} D.F. Geesaman, K. Saito and A.W. Thomas,
Ann. Rev. Nucl. Part. Sci. {\bf 45}, 337 (1995).
\bibitem{shadow2} M. Arneodo {\em et al.}, Nucl. Phys. B {\bf
441}, 3 (1995).
\bibitem{Kope96} B. Kopeliovich, Proc. Workshop Hirschegg '95,
ed. by H. Feldmeier and W. Norenberg, Darmstadt, 1995, 102,
hep-ph/9609385.
\bibitem{Brod98} S.J. Brodsky, A. Hebecker and E. Quack, Phys.
Rev. D {\bf 55}, 2584 (1997).
\bibitem{E772} D.M. Alde {\em et al.}, Phys. Rev. Lett. {\bf 64},
2479 (1990).
\bibitem{hawker} E. Hawker {\em et al.},  Phys. Rev. Lett. {\bf
80}, 3715 (1998).
\bibitem{esk} K.J. Eskola, V.J. Kolhinen and P.V. Ruuskanen,
Nucl. Phys. B {\bf 535}, 351
(1998); K.J. Eskola, V.J. Kolhinen and C.A. Salgado,
hep-ph/9807297.
\bibitem{MRST} A.D. Martin {\it et al}., Eur. Phys. J. C {\bf 4},
463 (1998).
\bibitem{CTEQ5} H.L. Lai {\it et al}., hep-ph/9903282.
\bibitem{NMC} M. Arneodo {\it et al}., Nucl. Phys. B {\bf 481},
3 (1996); Nucl. Phys. B {\bf 481}, 23 (1996).
\bibitem{Gale99} C. Gale, S. Jeon, and J. Kapusta, Phys. Rev.
Lett. {\bf 82}, 1636 (1999).
\bibitem{Marc99} E. Marco and E. Oset, Nucl. Phys. A {\bf 645},
303 (1999).
\bibitem{Brod99} Eq.\,\ref{eq:2} and Eq.\,7 of
\cite{brodsky} differ because we define
$\Delta x_1=\Delta E_{parton}/E_{beam}$.
Eq.\,7 of \cite{brodsky} uses
$\Delta x_1=\Delta E_{parton}/E_{parton}$.
S.J. Brodsky, private communication (1999).
\bibitem{Wang98} X.-N. Wang, Phys. Rev. Lett. {\bf 81},
2655 (1998).
\bibitem{baier3} Note that \cite{baier2} indicates that
the estimate 0.02 GeV/fm$^2$ given in \cite{baier1}
should be increased by a factor of 2.
\bibitem{rst99} R.S. Towell {\it et al}., to be submitted.

\end{references}
\end{document}